\documentclass[twocolumn]{aastex63}
\received{---}
\revised{---}
\accepted{---}
\submitjournal{ApJ}

\shorttitle{\textit{ASTROSAT} view of MAXI~J1803--298}
\shortauthors{Chand et al.}
\usepackage[flushleft]{threeparttable}
\usepackage{url}

\begin{document}

\title{\textit{AstroSat} view of the newly discovered X-ray transient MAXI~J1803--298 in the Hard-intermediate state}

\correspondingauthor{Parijat Thakur}
\email{parijat@associates.iucaa.in, parijatthakur@yahoo.com}

\author{Swadesh Chand}
\affil{Department of Pure and Applied Physics, Guru Ghasidas Vishwavidyalaya (A Central University), Bilaspur (C. G.)- 495009, India}

\author{G. C. Dewangan}
\affiliation{Inter-University Centre for Astronomy and Astrophysics, Post Bag 4, Ganeshkhind, Pune - 411007, India} 

\author{Parijat Thakur}
\affiliation{Department of Pure and Applied Physics, Guru Ghasidas Vishwavidyalaya (A Central University), Bilaspur (C. G.)- 495009, India}

\author{Prakash Tripathi}
\affiliation{Inter-University Centre for Astronomy and Astrophysics, Post Bag 4, Ganeshkhind, Pune - 411007, India}

\author{V. K. Agrawal}
\affiliation{Space Astronomy Group, ISITE Campus, U. R. Rao Satellite Center, Outer Ring Road, Marathahalli, Bangalore - 560037, India}




\begin{abstract}

We perform comprehensive temporal and spectral analysis of the newly discovered X-ray transient MAXI~J1803--298 using an AstroSat target of opportunity observation on May 11, 2021 during its outburst. The source was found to be in the hard intermediate state. We detect type C quasi-periodic oscillations (QPOs) at the frequencies of $\sim5.4$ Hz and $\sim6.3$ Hz along with a sub-harmonic at $\sim2.8$ Hz in the $3-15$ keV band. The frequency and fractional rms amplitude of the QPO in the $15-30$ keV band are found to be higher than those in the $3-15$ keV band. We find soft lags of $\sim3.8$ ms and $\sim6.8$ ms for the respective QPOs at $\sim5.4$ Hz and $\sim6.3$ Hz, whereas soft lag of $\sim4.7$ ms is found at the sub-harmonic frequency. The increase in the soft lags at the QPO frequencies with energy is also observed in other black hole transients and is attributed to the inclination dependence of the lags. The rms-energy spectra indicate the power-law component to be more variable than the disk and the reflection components. We find a broad iron line with an equivalent width of $\sim0.17-0.19$ keV and a reflection hump above $\sim12$ keV in the energy spectrum. Based on the X-ray spectroscopy and considering the distance to the source as 8 kpc, the estimated mass ($\sim8.5-16$ M$_\odot$) and spin ($a\gtrsim0.7$) of the black hole suggest that the source is likely to be a stellar mass Kerr black hole X-ray binary.

\end{abstract}



\section{Introduction} \label{sec:intro}

Black hole X-ray transients (BHTs) are accreting systems that show intermittent outbursts after a prolonged period of quiescence due to abrupt changes in their mass accretion rate. Sudden increase in the luminosity by several orders of magnitude during such outburst lead to the discovery of these kind of systems \citep{Tanaka and Shibazaki1996, Shidatsu et al.2014, Plant et al.2015}. The evolution of an usual outburst takes place through the state transition of the 
BHTs from the low/hard state (LHS) to the high/soft state (HSS) via the hard and soft intermediate states \citep[HIMS and SIMS;][]{Belloni et al.2005, Belloni2010}. These states can be identified on the hardness intensity diagram \citep[HID;][]{Belloni et al.2005, Homan and Belloni2005, Gierlinski and Newton2006, Fender et al.2009, Belloni2010} based on the definite spectral and timing properties, exhibited by the BHTs in the respective states. The HSS is mainly characterized by the thermal emission from an optically thick and geometrically thin accretion disk and associated with a weak fractional root-mean-square (rms) variability of few percent. In this state, the accretion disk is usually extended close to the inner-most stable circular orbit \citep[ISCO;][]{Gierlinski and Done2004, Steiner et al.2010, Belloni et al.2011}. Contrarily, the LHS is dominated by the hard power-law emission, originated through the Comptonization process, having a  high energy cutoff at $\sim100$ keV. The power density spectra (PDS) in this state is associated with a strong fractional rms variability of $\sim30\%$ \citep{McClintock and Remillard2006, Belloni et al.2011, Zhou et al.2013, Shidatsu et al.2014, Ingram et al.2017}. The true extent of the inner accretion disk in the LHS is still a matter of debate. \citet{ Esin et al.1997} proposed that the optically thick and geometrically thin accretion disk is replaced by a hot advection-dominated accretion flow (ADAF) in the LHS, and thus the accretion disk appears to be truncated away from the ISCO \citep[see][]{McClintock et al.1995, McClintock et al.2001, McClintock et al.2003, Narayan and Yi1995, Narayan et al.1996, Esin et al.2001, 
Ingram et al.2009, Plant et al.2015, Furst et al.2015, Garcia et al.2015, Ingram et al.2016}. However, the controversy still remains as it is suggested by some of the previous results that the inner accretion disk may appear to be truncated due to the photon pile-up issue present in the data \citep{Done and Trigo2010, Miller et al.2010}. The inner disk radius can be estimated either by modeling the thermal emission from the disk \citep{Zhang et al.1997} or the fluorescent iron emission line, arising due to the reflection of the hard coronal X-rays from the accretion disk \citep{Fabian et al.1989}. However, the properties of the BHTs in the intermediate states are not clear enough till now, and the energy spectrum is found to be associated with both power-law component and contribution from the accretion disk.

The transition of BHTs from one state to another are also associated with certain changes in the timing properties of such systems. Among these, the appearance of the low-frequency quasi-periodic oscillations (LFQPOs) at the frequency range of $0.05-30$ Hz are very common in the intermediate states and also in the LHS. The LFQPOs are classified as type A, B and C depending on certain characteristics exhibited by them. Type C QPOs generally appear in the HIMS and the LHS as narrow and variable peaks at the frequency ranging from few mHz to several Hz and depict strong fractional rms ($\geqslant10\%$). On the other hand, Type B QPOs are found in the SIMS at the frequency range of $4-6$ Hz and show relatively weaker fractional rms 
($\sim4\%$). Type A QPOs are rarely seen and appears as broad and weak peaks (with few percent of fractional rms) in the frequency range of $6-8$ Hz \citep{Wijnands and van der Klis1999, Remillard et al.2002, Casella Belloni and Stella2005, Homan and Belloni2005, 
Motta et al.2011, Motta2016, Stevens and Uttley2016, Gao et al.2017, Stevens et al.2018, Belloni et al.2020}. The study of the LFQPOs and the change in their properties such as time-lags and fractional rms amplitude with energy allow to understand the mechanism of the variability occurring in the timescale of seconds to days, and also provide a connection to the spectral component, responsible for the origin of such energy-dependent variability in BHTs. 

MAXI J1803--298 is a newly detected bright uncatalogued X-ray transient discovered by the Monitor of All-sky X-Ray Image 
\citep[MAXI;][]{Matsuoka et al.2009} with $4-10$ keV flux level of $46\pm18$ mcrab on May 01, 2021. The source was found to be located at  RA $=$ 18$^{h}$3$^{m}$41$^{s}$ and Dec. $=$ -29$^\circ$48$'$14$"$, estimated with an elliptical error region of long and short radii of 0.31$^\circ$  and 0.26$^\circ$, respectively \citep[$90\%$ confidence limit;][]{Serino et al.2021}. Following this, Swift/XRT follow-up observation detected the source with the $0.2-10$ keV flux level of $7.51\times10^{-10}$ erg cm$^{-2}$ s$^{-1}$, and also the optical counterpart of the transient system \citep{Gropp et al.2021, Hosokawa et al.2021}. The optical spectroscopy, carried out with Southern African Large Telescope (SALT) on May 2, 2021, suggested the source to be a low-mass black hole X-ray binary system \citep{Buckley et al.2021}. The radio counterpart was also detected by \citet{Espinasse et al.2021} on May 4, 2021. Further follow-up observations of the source with several other telescopes in the X-ray band reveal the detection of LFQPOs and iron line in the source. Moreover, the follow-up observation by MAXI/GSC on May 12, 2021 indicated the hard-to-soft transition of the source, where the spectral parameters indicate the system likely to be a black hole X-ray binary \citep{Shidatsu et al.2021}. In this work, we perform a detailed spectral and temporal analysis of the newly discovered X-ray transient source MAXI~J1803--298 using an AstroSat target of opportunity (ToO) observation. We detect LFQPOs, which exhibit similar nature as type C QPOs. We have also studied the energy dependent variability of the QPO properties. Moreover, we find the presence of a broad iron line, when the source was in the HIMS. The spectral modeling of the time-averaged X-ray energy spectra is performed with three model combinations including the blurred reflection model. The remainder of the paper is organized as follows. We
present the observation and data reduction in Section 2, and Section 3 contains the analysis and results of our temporal and spectral study. Finally, Section 4 is devoted to discussion and concluding remarks.

\section{Observation and Data Reduction} \label{sec:style}

We proposed a ToO observation of the X-ray transient MAXI~J1803--298 with India's first multi-wavelength satellite AstroSat. The follow-up observation 
(ObsID: T04\_003T01\_9000004368) of the source was carried out by the AstroSat on May 11 at 01:09 UTC (MJD 59345) and lasted up to May 11 at 11:59 UTC.

We obtained the level2 data of Soft X-Ray Telescope \citep[SXT;][]{Singh et al.2016, Singh et al.2017} from the ISSDC website\footnote {\url{https://astrobrowse.issdc.gov.in/astro_archive/archive/Home.jsp}}. The data from individual orbits were merged using the SXT event merger tool\footnote{\url{https://www.tifr.res.in/~astrosat_sxt/dataanalysis.html}} to produce an exposure-corrected merged clean event file. Using this merged clean file, we first extracted the source count rate from a circular region of 15$^{'}$ radius centered at the source position using the XSELECT V2.6d tool. The source count rate was significantly larger than the threshold value for pile-up (i.e. $>40$ cts/s) in the photon counting (PC) mode\footnote{\url{https://www.tifr.res.in/~astrosat_sxt/instrument.html}}, which gives hint about the presence of possible pile up in the energy spectrum. To mitigate the pile up effect, we excluded the source counts from the center region of the point spread function (PSF) until the source count rate became below the threshold rate for the pile-up. We found that for an annulus region with an inner radius of 7.5$^{'}$ and outer radius of 15$^{'}$, the source count rate becomes lower than the above mentioned threshold value. We, therefore, used this aperture size to extract the source spectrum. The large PSF (full-width at half maximum $\sim 2\arcmin$) and half power diameter (HPD $\sim 10\arcmin$) of the SXT, and the calibration sources present at the corners of the CCD do not leave the source free regions to extract the background spectrum. Because of this, we used the blank sky background file (SkyBkg$\_$comb$\_$EL3p5$\_$Cl$\_$Rd16p0$\_$v01.pha) provided by the SXT payload operation centre (POC). We also used the response matrix file (RMF) ``sxt$\_$pc$\_$mat$\_$g0to12.rmf", provided by the SXT POC. The SXT off-axis auxiliary response file (ARF) in accordance to the location of the source on the CCD was derived using the sxtARFModule tool\footnote{\url{https://www.tifr.res.in/~astrosat_sxt/dataanalysis.html}}. After the launch of the AstroSat, the gain of the on-board SXT instrument is shifted and correction corresponding to this gain shift has not been incorporated in the the current version of
the SXTPIPELINE. Hence, as per the recommendation by the SXT team, we corrected the gain shift in the SXT spectrum within Interactive Spectral Interpretation System \citep[ISIS v.1.6.2-40;][]{Houck and Denicola2000} by considering the slope fixed at unity and making the offset variable.

We also used the data observed by the Large Area X-Ray Proportional Counter \citep[LAXPC;][]{Yadav et al.2016a, Yadav et al.2016b, Agrawal et al.2017, Antia et al.2017}. We processed the LAXPC level1 data of the source using the LAXPC software (Laxpcsoft\footnote{\url{http://astrosat-ssc.iucaa.in/?q=laxpcData}}), and obtained the level2 data. We generated the lightcurves and energy spectrum using the standard task available within the Laxpcsoft\footnote{\url{https://www.tifr.res.in/~astrosat_laxpc/LaxpcSoft.html}}. In this work, we have used only the LAXPC20 data for the spectral and timing analysis. We discarded the LAXPC10 data due to its low gain and response, and LAXPC30 data as it is currently not in the working condition\footnote{\url{http://astrosat-ssc.iucaa.in/}}.

\section{Analysis and Results} \label{sec:floats}

\subsection{Timing Analysis}

We extracted the source and background lightcurves from the LAXPC20 data in the $3-60$ keV band. In the upper panel of Figure 1, we show the background-subtracted source lightcurve with a time bin size of 10 s derived in the same energy band. The source intensity remains almost constant over the entire observation period with an average count rate of $\sim663$ cts/s. Also, the hardness ratio (HR), derived between $3-7$ keV and $7-15$ keV bands as shown in the lower panel of Figure~1, does not indicate any significant variation in the lightcurve throughout this particular observation.

\begin{figure*}[ht!]
\plotone{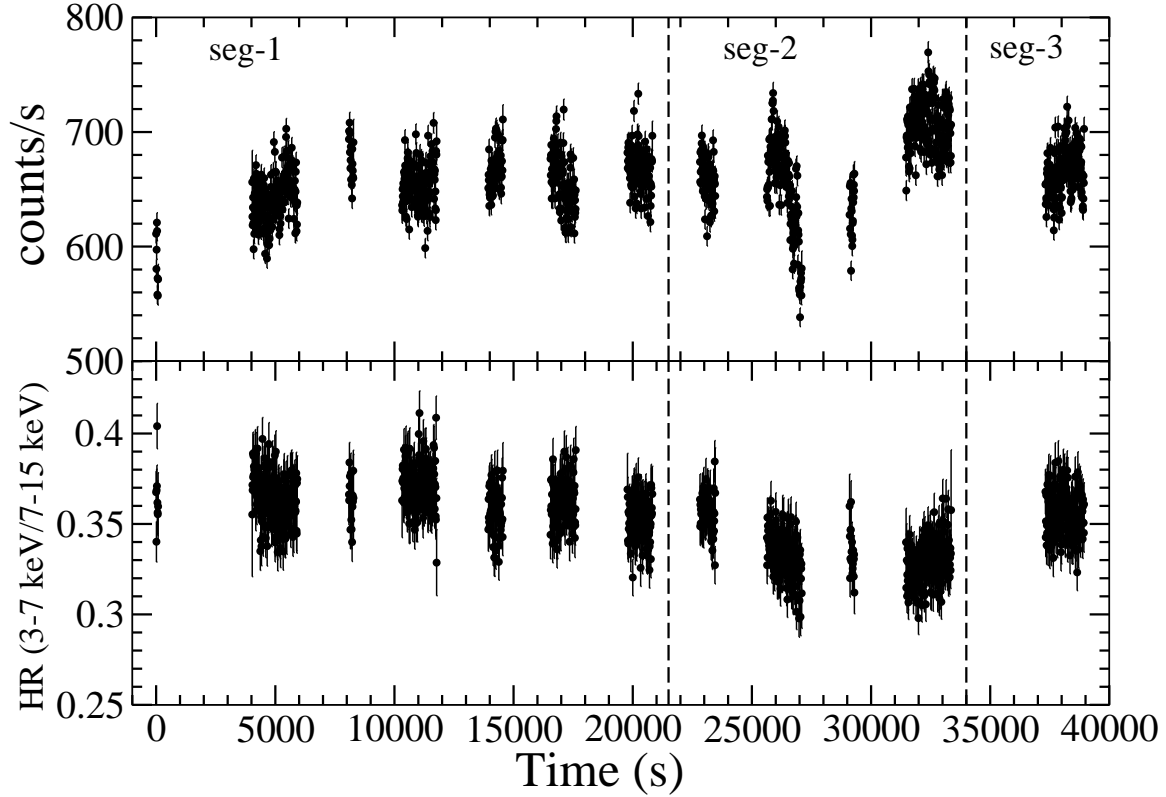}
\caption{Background-subtracted lightcurve with the time bin size of 10 s derived in the $3-60$ keV band using the LAXPC20 data (upper panel) and the hardness ratio derived between $3-7$ keV and $7-15$ keV bands (lower panel).}
\end{figure*}

We derived PDSs in the $3-15$ keV and $15-30$ keV bands using lightcurves binned with 0.02 s corresponding to the Nyquist frequency of 25 Hz. We did not use the poor quality data above 30 keV. We considered intervals 
of 16384 bins and obtained rms normalized PDS in each interval by dividing the Leahy normalized power spectra by an average rate \citep{Leahy et al.1983, Belloni and Hasinger1990}. We computed the final PDS by averaging all the rms normalized PDSs for different intervals and propagated the error bars appropriately. We rebinned the final PDS in the frequency space using a geometric binning factor of 1.05. Further, we subtracted the dead-time corrected Poisson noise level from each PDS \citep[see][]{Zhang et al.1995, Yadav et al.2016b, Agrawal et al.2018}. Since, LAXPC has a dead-time of $43\mu$s that can affect the true source rms, we followed the prescription given by \citet{Bachetti et al.2015} in order to correct the effect of dead-time on the source rms using the formula 
rms$_{\rm{in}}=\rm{rms_{det}} (1+\tau_d r_{\rm{det}}/N)$. Here, rms$_{\rm{in}}$ is the dead-time corrected rms, 
rms$_{\rm{det}}$ is the instrument detected rms, $\tau_d$ is the dead-time, $r_{\rm{det}}$ is the detected count rate and $N$ is the number of proportional counter units \citep{van der klis1988, Zhang et al.1995, Sreehari et al.2020}. In addition, all the PDSs are corrected for the corresponding background rates \citep{Yadav et al.2016b, Rawat et al.2019}.
Figure 2 shows the PDSs derived using the full LAXPC exposure time in the $3-15$ keV and $15-30$ keV bands in the left and right panels, respectively. We modeled the PDS in $3-15$ keV band with a power-law and five Lorentzians, and detected a LFQPO at $\sim5.4$ Hz along with a sub-harmonic at $\sim2.8$ Hz with the detection significance of 16$\sigma$ and 12.5$\sigma$, respectively.
However, it is to note here that since we could not fit the broad QPO feature at $\sim5.4$ Hz with a single Lorentzian component, an additional Lorentzian component centered at $\sim6.3$ Hz was used. The detection significance level of the QPO at $\sim6.3$ Hz is found to be $6.8\sigma$, which is relatively lower than that of the QPO at $\sim5.4$ Hz. On the other hand, the PDS in the $15-30$ keV band is fitted with only two Lorentzian components. In this band, we also detected the QPO at the frequency of $\sim5.5$ Hz with a significance of 16.4$\sigma$. It appears that the QPO in this band is found to be shifted to a slightly higher frequency than that of the $3-15$ keV band. However, the sub-harmonic was absent in this band. The best-fit parameters are listed in Table 1, where all the errors are quoted in the 90\% confidence level. In the $3-15$ keV band, the quality factors ($Q$=$\nu_{\rm{centroid}}$/FWHM; FWHM being the full width at half maximum) are $\sim7.3$ (for the QPO at $\sim5.4$ Hz) and $\sim4.3$ (for the QPO at $\sim6.3$ Hz), whereas it is found to be $\sim2.1$ for the sub-harmonic. Moreover, the estimated Q-value of the QPO at $\sim5.5$ Hz in the $15-30$ keV band is $\sim 3.3$, which is comparatively lower than that found in the $3-15$ keV. The fractional rms amplitudes are $\sim9\%$ (for the QPO at $\sim5.4$ Hz) and 
$\sim7.3\%$ (for the QPO at $\sim6.3$ Hz), whereas it is $\sim7\%$ for the sub-harmonic. However, the fractional rms variability amplitude of the QPO in the $15-30$ keV band is $\sim20\%$, which appears to be higher than that obtained in the $3-15$ keV band.

Moreover, to check the time-dependent behavior of the QPO, we derived the dynamic power spectrum (DPS) in the frequency range of $0.1-10$ Hz for both the $3-15$ keV and $15-30$ keV bands using the LAXPC tool `$\rm{laxpc\_dynpower}$\footnote{\url{http://astrosat-ssc.iucaa.in/laxpcData}}' over the full exposure time of LAXPC20 data \citep{Jithesh et al.2019, Rawat et al.2019}. The DPS derived in the $3-15$ keV band is depicted in Figure 3, where a clear change in the QPO frequency in the interval of $\sim 5.4-6.4$ Hz can be seen over the time. Thus, we divided the $3-15$ keV band lightcurve into three time segments: (i) seg-1: upto $21.5$ ks, (ii) seg-2: $21.5-34$ ks and (iii) seg-3: $34-40$ ks (see Figure~1). Following this, we derived rms normalized and Poisson noise (dead-time corrected) subtracted PDS for each segment.
The QPO frequency is found at $\sim 5.4$ Hz in the seg--1, then increases to $\sim6.4$ Hz in the seg--2, and finally moves down 
to $\sim 5.5$ Hz in the seg--3. However, no significant systematic variation in the fractional rms amplitude of the QPO between these three segments is noticed. The segment-wise PDSs are shown Figure 4 and the corresponding best-fit parameters are listed in Table 2. No change in the frequency of the QPO is observed in the $15-30$ keV band.

In addition to the above, we investigated energy dependence of the fractional rms amplitude variability of both the QPOs and the sub-harmonic using the full LAXPC exposure time. In this regard, we divided the $3-30$ keV band into seven energy bands (i.e. $3-5$, $5-7$, $7-10$, $10-15$, $15-20$, $20-25$, and $25-30$ keV), and derived the fractional rms variability at the QPOs and sub-harmonic frequencies from the PDSs of each energy band. The variation in fractional rms variability as a function of energy is depicted in Figure~5. It is clear that the fractional rms amplitudes of both the QPOs first increase from the disk emission dominated band ($3-5$ keV) to the power-law dominated band ($10-15$ keV), and then either decrease or become flat most likely due to the presence of a less variable reflection continuum (see Figure~5). A similar trend is found in the rms-energy spectrum of the sub-harmonic, where the amplitude of the variability is slightly lower than those of the QPOs. It is to note here that the error bars in the last few data points at the high energy bands are very large for the rms-energy spectra of both the QPOs and the sub-harmonic.

\begin{figure*}
\plottwo{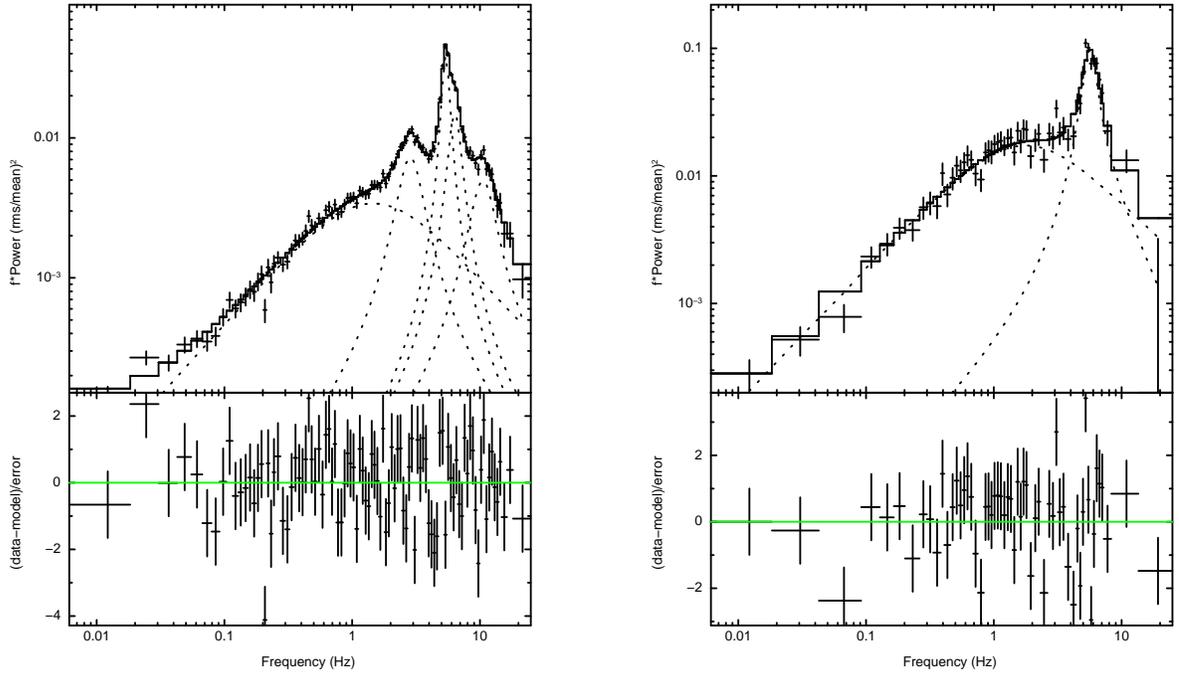}{fig_3.eps}
\caption{Power density spectra derived in the $3-15$ keV band (left panel) and $15-30$ keV band (right panel). Filled black circles represent the data, and the red solid lines indicate the model.\label{fig:f2}}
\end{figure*}

\begin{figure*}[ht!]
\plotone{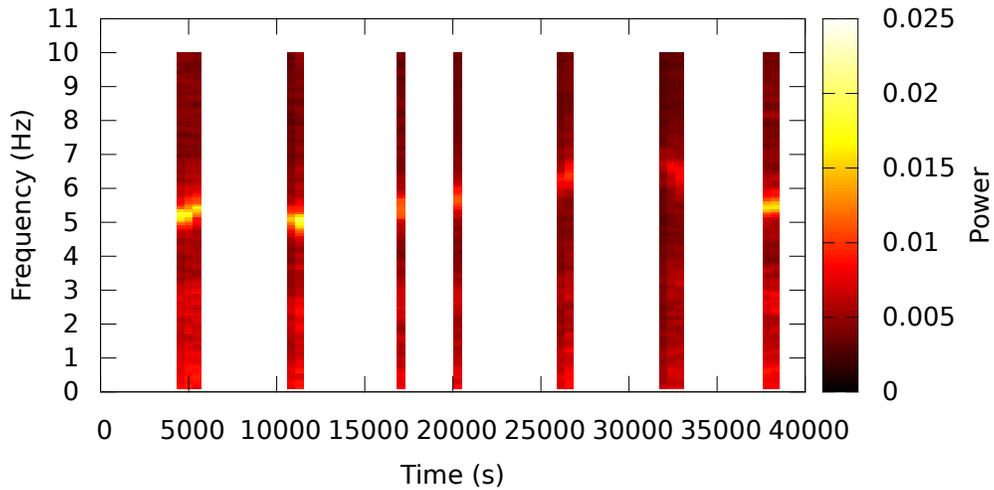}
\caption{Dynamic power spectrum, derived in the $3-15$ keV band, shows a clear change in the QPO frequency during the observation. The observation starts on MJD 59345.}
\end{figure*}

\begin{figure*}
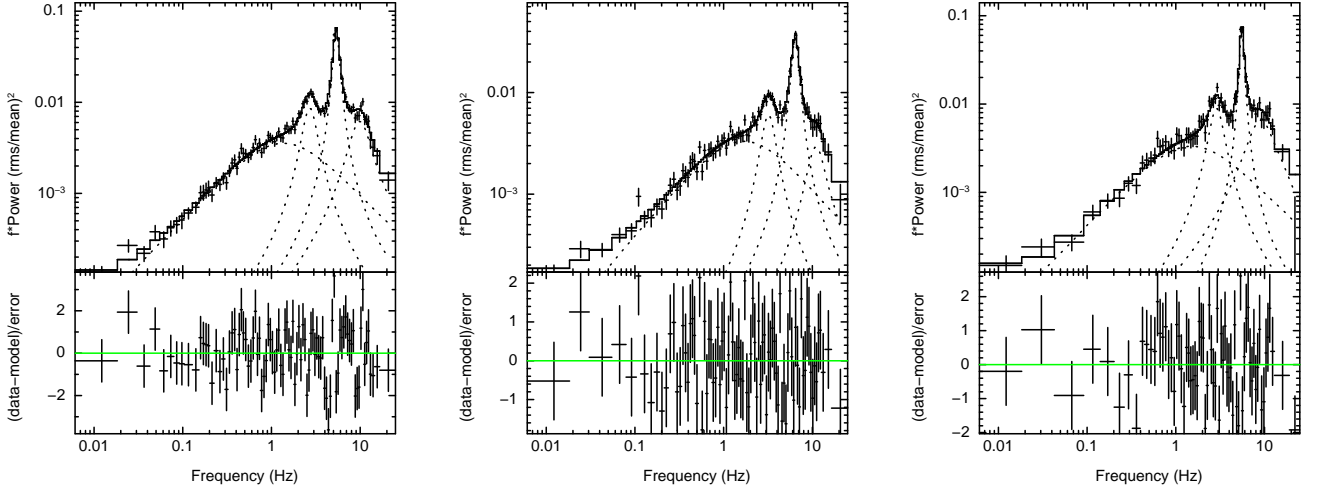

\gridline{\fig{fig_16.eps}{0.315\textwidth}{}
\fig{fig_17.eps}{0.315\textwidth}{}
\fig{fig_18.eps}{0.315\textwidth}{}
                    } 
\caption{Power Density spectra derived in the $3-15$ keV band for the seg--1 (left panel), seg--2 (middle panel) and seg--3 (right panel).\label{fig:pyramid}}
\end{figure*}

To investigate the nature of the variability, we computed time-lags at the QPOs and sub-harmonic frequencies by performing the cross-spectral analysis \citep{Vaughan and Nowak1997, Nowak et al.1999} over the full LAXPC exposure time. For two evenly sampled time series $x(t)$ and $y(t)$, the complex-valued Fourier transforms $X(f)$ and $Y(f)$ were calculated, where $t$ is the time bin in seconds and $f$ is the Fourier frequency in Hz. From these, the Cross-spectrum as $C_{X,Y}(f) = X^{*}(f)Y(f) = |X(f)||Y(f)|\exp(-i(\phi_{X(f)}-\phi_{Y(f)}))$ was calculated, where $X^{*}(f)$ is the complex conjugate of $X(f)$, and $\phi_{X(f)}$ and $\phi_{Y(f)}$ are the phase angles. After this, the cross-spectrum was averaged over $M$ non-overlapping lightcurves segments and $K$ frequency bins. Using the averaged cross-spectrum $\overline{C}_{X,Y}(f)$,  the phase lags were obtained as $\phi(f) = \rm{arg}(\overline{C}_{X,Y}(f))$, and it was then transformed into the time-lags as $\tau(f) = \phi(f)/2\pi{f}$. The error on the time lag was calculated using the prescription given in \cite{Nowak et al.1999}. To derive the time-lags from the LAXPC20 data, we used the LAXPC subroutine `laxpc\_find\_freqlag', which uses the method described in \citet{Vaughan and Nowak1997} and \citet{Nowak et al.1999}. In order to study the time-lags as a function of frequency, 
two lightcurves were extracted in the $3-5$ keV and $9-12$ keV bands, which are dominated by thermal emission of the accretion disk and X-ray power-law continuum, respectively. Each lightcurve was then divided into 4682 segments of length 3.2 ks, and the Fourier frequency was binned to an interval of 0.28 Hz.
The variation of time-lags as a function of Fourier frequency is shown in Figure~6, where the negative lag represents the soft lag. This suggests that the photons from the soft energy band lags behind the ones from the hard energy band. We have marked the time-lags at the QPOs and the sub-harmonic frequencies by the three dotted vertical lines in Figure~6, where the observed soft lag is $3.8\pm1.0$ ms (for the QPO at $\sim5.4$ Hz), $6.8\pm1.9$ ms (for the QPO at $\sim6.3$ Hz) and $4.7\pm4.2$ ms (for the sub-harmonic at $2.8$ Hz). 

In addition to the above, we have also investigated the energy dependence of the time-lags observed at the QPOs and the sub-harmonic frequencies across the $3-30$ keV band. Here, the same energy bands, employed above in deriving the fractional rms variability, were used to calculate the time-lags. We considered the $3-5$ keV band as the reference band and binned the Fourier frequency at an interval of 0.3 Hz. The energy dependent time-lags estimated at the QPOs and sub-harmonic frequencies are given in Figure~7. We observed soft-lag at both the QPOs and the sub-harmonic frequencies, and found that the soft-lag for both the QPOs first increases (negatively) upto $\sim12$ keV, and then almost becomes flat or start decreasing. The error bars are large at the higher energy bands. However, the trend of the lag-energy spectra is found to be similar to the rms-energy spectra of the QPOs, where both the soft-lag and fractional rms increase with energy upto $\sim12$ keV, become apparently almost flat or start decreasing.
A similar trend between the time-lags and fractional rms variability has also been found for the sub-harmonic. However, a mismatch in the last data point of the time-lags and rms spectra is present with large error bars.

\begin{figure*}
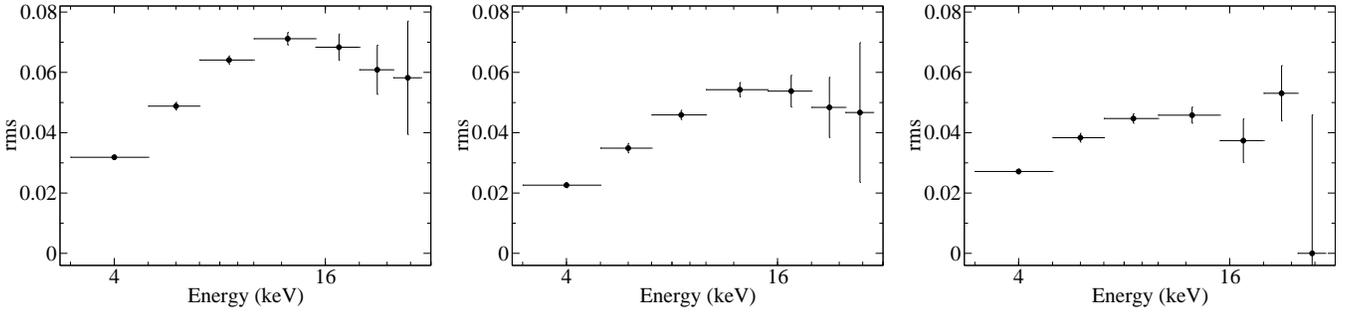

\gridline{\fig{fig_4.eps}{0.315\textwidth}{}
\fig{fig_5.eps}{0.315\textwidth}{}
\fig{fig_6.eps}{0.315\textwidth}{}
                    } 
\caption{The fractional rms variability as a function of energy for the QPO at $\sim5.4$Hz (left panel), $\sim6.3$ Hz (middle panel) and the sub-harmonic at $\sim2.8$ Hz (right panel).\label{fig:pyramid}}
\end{figure*}

\begin{figure*}
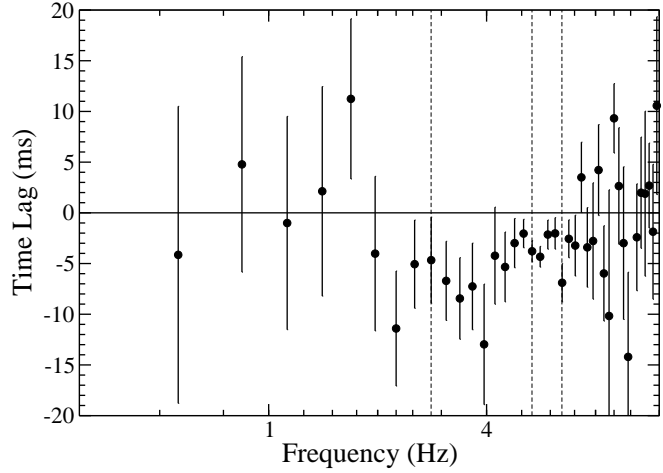

\gridline{\fig{fig_7.eps}{0.48\textwidth}{}
                } 
\caption{Time-lags as a function of frequency estimated between the $3-5$ keV and $9-12$ keV bands. The vertical dotted lines show the position of the QPOs and the sub-harmonic, where soft or negative lags have been detected.}
\end{figure*}


\begin{figure*}
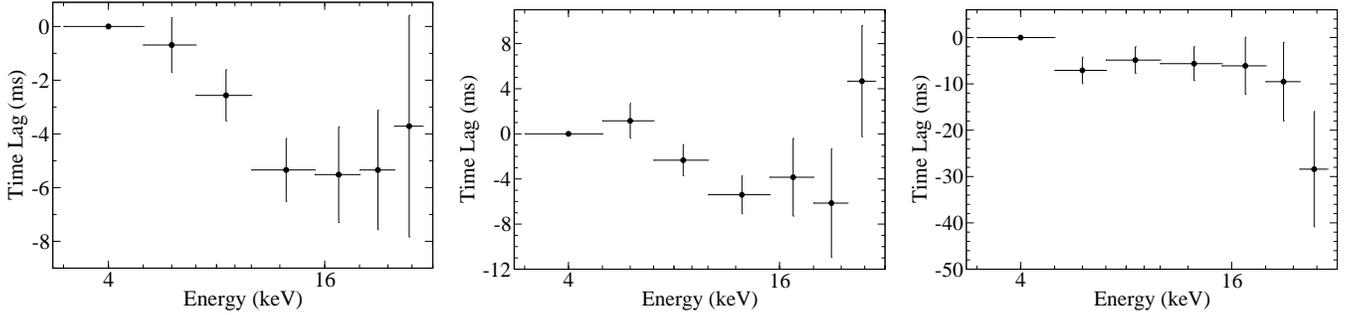

\gridline{
        \fig{fig_8.eps}{0.315\textwidth}{}
        \fig{fig_9.eps}{0.315\textwidth}{}
        \fig{fig_10.eps}{0.315\textwidth}{}
                }
\caption{Time-lags derived as a function of energy for the QPO at $\sim5.4$Hz (left panel), $\sim6.3$ Hz (middle panel) and the sub-harmonic at $\sim2.8$ Hz (right panel).}
\end{figure*}

\begin{table*}[!ht]
	\caption{Best-fit QPOs and sub-harmonic parameters obtained from the PDS in the $3-15$ keV and $15-30$ keV bands.}
   \centering
   \begin{tabular}{l@{\hskip 0.8in}c@{\hskip 0.8in}c@{\hskip 0.8in}c} 
      \hline
      \hline
    Component & Parameter & 3--15 keV & 15--30 keV\\
     \hline
     QPO1 & $\nu$ (Hz) & 5.38$\pm$0.02 & 5.6$\pm0.1$ \\ 
     & $\rm{FWHM}$ (Hz) & 0.73$\pm$0.07 & 1.7$\pm0.2$ \\ 
     & $Q$ & 7.3$^{+0.6}_{-0.5}$ & 3.3$\pm0.4$ \\ 
     & $\rm{rms}$ [\%] & 9$\pm3$ & 20.1$\pm6.4$ \\ 
     QPO2 & $\nu$ (Hz) & 6.3$\pm$0.1 & \nodata \\ 
     & $\rm{FWHM}$ (Hz) & 1.4$\pm$0.2 & \nodata \\
     & $Q$ & 4.3$^{+0.8}_{-0.5}$ & \nodata \\
     & $\rm{rms}$ [\%] & 7.3$^{+3.6}_{-3.3}$ & \nodata \\
     Sub-harmonic & $\nu$(Hz) & 2.77$\pm0.04$ & \nodata \\ 
     & $\rm{FWHM}$ (Hz) & 1.29$^{+0.16}_{-0.15}$ & \nodata \\ 
     & $Q$ & 2.1$\pm0.2$ & \nodata  \\
     & $\rm{rms}$ [\%] & 7.0$\pm2.5$ & \nodata \\ 
     & $\chi^2$/dof & 125.3/82 & 89.4/53 \\
     \hline
   \end{tabular}
   \begin{tablenotes}
   \item \textbf{Notes.} $\nu$ -- Centroid frequency, FWHM -- Width, $Q$ -- Quality factor, rms -- Fractional root-mean-square variability.	
   \end{tablenotes}
\end{table*}

\begin{table*}[!ht]
	\caption{Best-fit QPOs and sub-harmonic parameters obtained from the PDSs derived in the three segments in the $3-15$ keV band.}
   \centering
   \begin{tabular}{l@{\hskip 0.2in}c@{\hskip 0.2in}c@{\hskip 0.2in}c@{\hskip 0.2in}c@{\hskip 0.2in}c@{\hskip 0.2in}c@{\hskip 0.2in}c} 
      \hline
      \hline
    Segment & $\nu_{\rm{qpo}}^1$(Hz) & $\rm{FWHM_{qpo}^2}$(Hz) & $\rm{rms_{qpo}^3}$[\%] & $\rm{\nu_{har}^4}$(Hz) & $\rm{FWHM_{har}^5}$(Hz) & $\rm{rms_{har}^6}$[\%] & $\chi^2$/dof\\
     \hline
     1 & 5.36$\pm0.01$ & 0.84$\pm0.04$ & 12.4$^{+2.3}_{-2.4}$ & 2.64$\pm0.05$ & 1.07$^{+0.13}_{-0.15}$ & 7.2$\pm2.7$ & 136.1/79\\
     2 & 6.37$\pm0.03$ & 1.2$\pm0.1$ & 10.1$^{+2.5}_{-2.6}$ & 3.14$\pm0.08$ & 1.2$\pm0.3$ & 6$\pm3$   & 77.6/73\\
     3 & 5.54$\pm0.02$ & 0.52$\pm0.07$ & 11$\pm3$ & 2.8$\pm0.1$ & 1.1$\pm0.3$ & 7$\pm4$ & 69.2/55\\
     \hline
   \end{tabular}
  \begin{tablenotes}
  \item \textbf{Notes.} $^{1,2,3}$Centroid frequency, width and fractional rms of QPO,
  $^{4,5,6}$Centroid frequency, width and fractional rms of sub-harmonic.
    \end{tablenotes}
\end{table*}

\subsection{Broadband Spectral Analysis}

We carried out the broadband X-ray spectral analysis by simultaneously fitting the SXT ($0.7-6$ keV) and LAXPC20 ($3-50$ keV) spectral data in the $0.7-50$ keV band. All the spectral modeling were performed using Interactive Spectral Interpretation System \citep[ISIS, v.1.6.2-40;][]{Houck and Denicola2000}, and the uncertainties are quoted at $90\%$ confidence level. We grouped the SXT spectral data to a minimum S/N of 5 and a minimum of 3 channels per bin, whereas the LAXPC20 spectral data was grouped to a minimum S/N of 3 and a minimum of 1 channel per bin. Initially, we fitted both the spectra simultaneously with a power-law component modified by the Galactic absorption model TBabs, considering the abundances and cross sections of the inter-stellar medium from \citet{Wilms et al.2000} and \citet{Verner et al.1996}, respectively. A constant component was also multiplied to the model to take care of the relative normalization between the two different instruments. We fixed the constant factor to 1 for the SXT spectral data and kept it free to vary for the LAXPC20 spectral data. Moreover, we added the multi-color disk blackbody model \citep[{\tt diskbb};][]{Mitsuda et al.1984} to address the spectrum originating from the accretion disk, and also incorporated $3\%$ systematic error to both the SXT and LAXPC20 spectral data to account for the calibration uncertainty as suggested by the instrumentation team\footnote{\url{https://www.tifr.res.in/~astrosat_sxt/dataanalysis.html}}.
However, this model setup {\tt tbabs*(powerlaw+diskbb)} left with significant residuals in the $5-9$ keV and above $\sim12$ keV bands implying the presence of the iron line and the reflection hump, respectively (see Figure 8). Hence, we have incorporated a {\tt gaussian} component for the iron line and convolved disk spectrum from the diskbb model with the {\tt thcomp} model \citep{Zdziarski et al.2020} to account for the Comptonization of the seed photons by the thermal electrons. Since the {\tt thcomp} model is a convolution model which redistributes a fraction of the seed photons to higher energies, the sampled energy range was extended to 1000 keV. The model 
{\tt constant*tbabs*(thcomp*diskbb+gaussian)} (hereafter Model-1) provides an admissible fit with $\chi^2/$dof $= 338/239$. The best fit spectral parameters are listed in Table 3, whereas the best-fit model is shown in Figure 9. We found the absorption column density ($N_H$) to be $\sim0.2\times 10^{22}$ cm$^{-2}$. The inner disk temperature ($kT_{\rm{in}}$) and the photon index ($\Gamma$) are found to be $\sim0.8$ keV and $\sim2.2$, respectively. We could not constrain the value of electron temperature ($kT_e$) and found only the lowest value to be $23$ keV, whereas the covering fraction($cov_{frac}$) is found to be 
$\sim0.4$. Furthermore, we kept the centroid line energy of the iron line fixed at $6.5$ keV and varied the line width along with the normalization. The best-fit value of the line width of $\sim1.2$ keV, whereas the estimated equivalent width (EW) of the iron line is found to be $0.17\pm0.02$ KeV.

\begin{table*}
	\caption{Best-fit broadband X-ray spectral parameters derived using Model 1--3. }
   \centering
   \begin{tabular}{l@{\hskip 0.5in}c@{\hskip 0.5in}c@{\hskip 0.5in}c@{\hskip 0.5in}c} 
      \hline
      \hline
	Component & Parameter & Model-1 & Model-2 & Model-3 \\
	\hline
	Constant & & 0.83 &  0.84 & 0.84 \\
	TBabs & $N_H^1$ ($10^{22}$ cm$^{-2})$ & 0.22$^\ddagger$ & 0.29$\pm0.01$ & 0.45$\pm0.01$ \\
	
		DISKBB & $kT_{in}^2$ (keV) & 0.82$^\ddagger$ & \nodata & \nodata \\
		& norm$^3$ & 1111.4$^{+5.3}_{-85.3}$ & \nodata & \nodata \\
		
	THCOMP & $\Gamma^4$ & 2.21$^{+0.03}_{-0.02}$ & 2.03$\pm0.03$ & 2.13$^{+0.04}_{-0.03}$ \\
			& $kT_e^5$ (keV) & $>23$ & 17$^{+18}_{-3}$ & $>95$ \\				
			& $cov_{frac}^6$ & 0.475$^{+0.004}_{-0.035}$ & 0.20$\pm0.02$ & $<0.14$ \\
			
	KERRBB  & $a^7$ & \nodata & $>0.95$ & 0.94$^{+0.03}_{-0.28}$ \\
			& $i^8$ ($^\circ$) & \nodata & 48$^{+4}_{-5}$ &  57$^{+17}_{-6}$\\
			& $M_{\rm{BH}}^9$ ($M_\odot$) & \nodata & 14$\pm2$ &  10.5$\pm2.0$\\
			& $\dot{M}^{10} (10^{18}$ g s$^{-1}$) & \nodata & 0.35$^{+0.08}_{-0.05}$ & 0.30$^{+0.27}_{-0.07}$ \\
			& $D^{11}$ (kpc) & \nodata & 8 (f) & 8 (f) \\
			
	GAUSSIAN & norm$^{12}$ ($10^{-3}$) & 20$\pm3$ & 21.5$\pm5.0$ & \nodata \\
			& $E_{\rm{line}}^{13}$ (keV) & 6.5 (f) & 6.5 (f) & \nodata \\
			& Width$^{14}$ (keV) & 1.2$\pm0.2$ & 1.4$\pm0.2$ & \nodata \\
			
	RELXILLCp & $q^{15}$ & \nodata & \nodata & $3$ (f) \\
		    & $R_{\rm{in}}^{16}$ ($R_{\rm{isco}}$) & \nodata & \nodata & 3.0$^{+2.2}_{-1.5}$ \\
		    & $log \xi^{17}$ & \nodata & \nodata & 4.0$^{+0.03}_{-0.10}$ \\
		    & $A_{\rm{Fe}}^{18}$ (solar) & \nodata & \nodata & 1(f) \\
		    & $R_{\rm{Ref}}^{19}$ & \nodata & \nodata & -1 (f) \\
		    & norm$^{20}$ ($\times10^{-3}$) & \nodata & \nodata & 21$^{+7}_{-4}$ \\
		    & $\chi^2$/dof & $\frac{338}{239}\approx1.4$ & $\frac{235.5}{237}\approx1$ & $\frac{256.6}{236}\approx1.1$ \\ 
			& Flux$_{0.7-50 \rm{keV}}$ (unabsorbed) & $1.3\times10^{-8}$ & $1.2\times10^{-8}$ & $1.4\times10^{-8}$ \\
	\hline 
    \end{tabular}
    \begin{tablenotes}
    \item \textbf{Notes.} $^1$Galactic absorption column density, $^2$ Inner disk temperature, $^4$Photon index, $^5$Electron temperature, $^6$Covering fraction, $^7$Spin of the black hole, $^8$Disk inclination, $^9$Mass of the black hole, $^{10}$Mass accretion rate in the unit of g s$^{-1}$, $^{11}$Distance from the source in the unit of kpc, $^{13}$\texttt{gaussian} line energy, $^{14}$\texttt{gaussian} line width, $^{15}$Emissivity index, $^{16}$Inner disk radius in the unit of $R_{\rm{isco}}$, $^{17}$Log of ionization parameter of the accretion disk, $^{18}$Iron abundance in the unit of Solar abundance, $^{19}$Reflection fraction, $^{3, 12 ,20}$ normalization parameter of the corresponding spectral parameter, $^\ddagger$ values of the errors are trivial and hence are not quoted, f-- indicates the fixed parameters, Flux is in the unit of erg cm$^{-2}$ s$^{-1}$.
    \end{tablenotes}
\end{table*}

\begin{figure}[!]
\centering
\includegraphics[width=8cm]{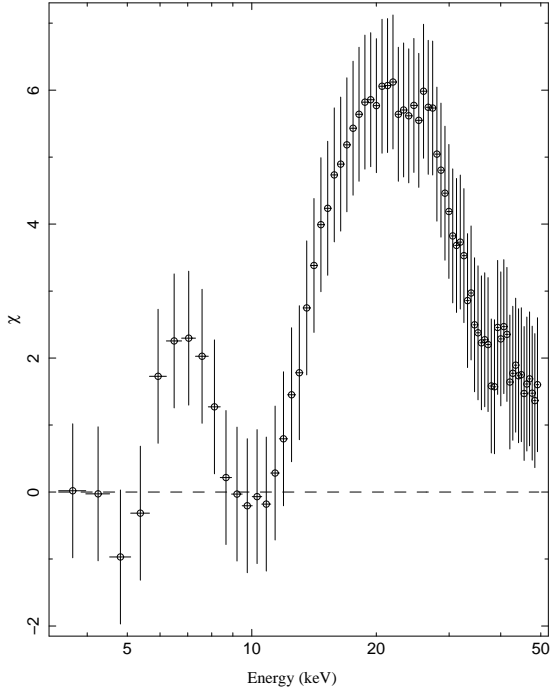}
\caption{Residual shows the deviations of the observed spectral data from the best-fitting model {\tt constant*tbabs(powerlaw+diskbb)}. The presence of the iron line excess in the $5-9$ keV region and the reflection hump above $\sim12$ keV region are clearly visible. Here, we have used only the LAXPC20 spectral data.}
\label{fig:xmm-newton}
\end{figure}

\begin{figure*}
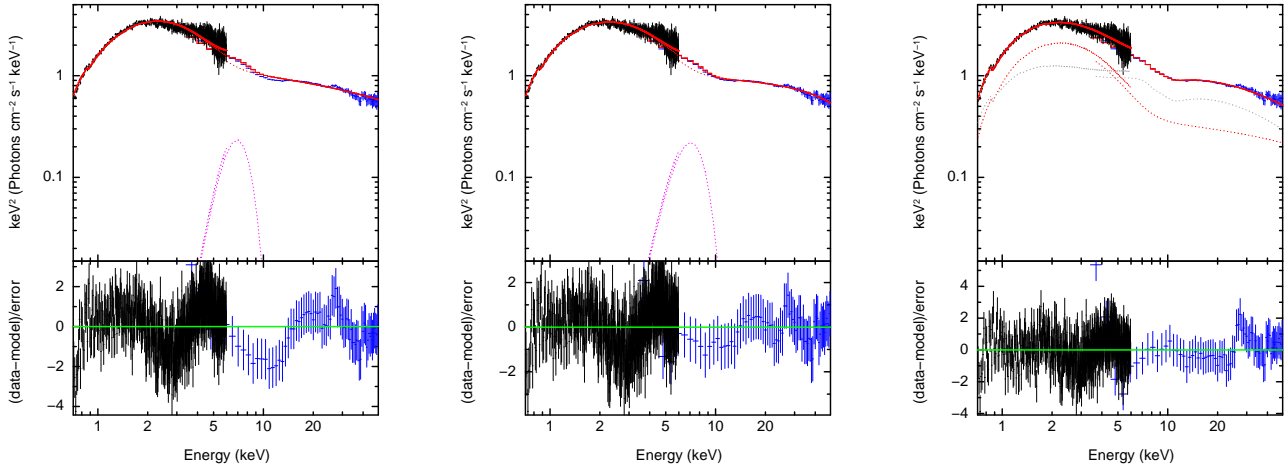

\gridline{\fig{fig_12.eps}{0.3\textwidth}{}
          \fig{fig_13.eps}{0.3\textwidth}{}
          \fig{fig_14.eps}{0.3\textwidth}{}
          }
\caption{Joint SXT and LAXPC20 spectral data fitted with Model-1 (left panel), Model-2 (middle panel) and Model-3 (right panel). Black circles are for the SXT spectral data, whereas the blue circles represent the LAXPC20 spectral data. The solid red lines indicate the extrapolated models of the continuum fit. The red dotted line represents the {\tt thcomp} component convolved with {\tt diskbb} (Model-1) and {\tt kerrbb} for (Model-2 and 3). The component {\tt gaussian} is represented by 
the dotted line in magenta. The {\tt relxillCp} component is shown by the light grey dotted line.
\label{fig:pyramid}}
\end{figure*}

\begin{table*}
	\caption{Best-fit segment wise X-ray spectral parameters derived using Model 2 and 3. }
   \centering
   \begin{tabular}{l@{\hskip 0.05in}c@{\hskip 0.05in}c@{\hskip 0.05in}c@{\hskip 0.2in}c@{\hskip 0.1in}r} 
      \hline
      \hline
      & & \hspace{0.6in} Seg 1 &  & \hspace{0.55in} Seg 2 \\
      \hline 
	Component & Parameter & Model-2 & Model-3 & Model-2 & Model-3\\
	\hline
	Constant & & 0.88 &  0.88 & 0.86 & 0.85 \\
	TBabs & $N_H^1$ ($10^{22}$ cm$^{-2})$ & 0.31$\pm0.02$ & 0.44$\pm0.02$ & 0.31$\pm0.02$ & 0.44$\pm0.02$ \\
		
	THCOMP & $\Gamma^4$  & 2.03$^{+0.03}_{-0.02}$ & 2.15$^{+0.04}_{-0.03}$ & 2.01$\pm0.03$ & 2.10$^{+0.06}_{-0.04}$ \\
			& $kT_e^5$ (keV) & 18$^{+19}_{-3}$ & $>83.5$ & 15$^{+7}_{-3}$ & $>44$\\				
			& $cov_{frac}^6$ & 0.20$^{+0.02}_{-0.01}$ & $0.2\pm0.1$ & 0.18$\pm0.01$ & $<0.2$\\
			
	KERRBB  & $a^7$ & $>0.9$ & 0.93$^{+0.05}_{-0.42}$ & $>0.9$ & 0.85$^{+0.07}_{-0.53}$\\
			& $i^8$ ($^\circ$) & 47$^{+8}_{-7}$ &  56.3$^{+8.5}_{-6.0}$ & 47$\pm7$ & 62$^{+9}_{-6}$\\
			& $M_{\rm{BH}}^9$ ($M_\odot$) & 15$\pm4$ &  13$^{+4}_{-3}$ & 14$^{+4}_{-3}$ & 9$^{+10}_{-3}$\\
			& $\dot{M}^{10} (10^{18}$ g s$^{-1}$) & 0.36$^{+0.10}_{-0.05}$ & 0.4$\pm0.1$ & 0.35$^{+0.08}_{-0.04}$ & 0.50$^{+0.80}_{-0.06}$ \\
			& $D^{11}$ (kpc) & 8 (f) & 8 (f) & 8(f) & 8 (f) \\
			
	GAUSSIAN & norm$^{12}$ ($10^{-3}$) & 20$\pm9$ & \nodata & 17$\pm5$ & \nodata \\
			& $E_{\rm{line}}^{13}$ (keV) & 6.5 (f) & \nodata & 6.5 (f) & \nodata \\
			& Width$^{14}$ (keV) & 1.3$^{+0.2}_{-0.3}$ & \nodata & 1.4 (f) & \nodata \\
			
	RELXILLCp & $q^{15}$ & \nodata & $3$ (f) & \nodata & 3(f) \\
		    & $R_{\rm{in}}^{16}$ ($R_{\rm{isco}}$) & \nodata & $<6.4$ & \nodata  & $<4.4$\\
		    & $log \xi^{17}$ & \nodata & 4.0$\pm0.1$ & \nodata & 4.0$^{+0.05}_{-0.06}$ \\
		    & $A_{\rm{Fe}}^{18}$ (solar) & \nodata & 1 (f) & \nodata & 1 (f) \\
		    & $R_{\rm{Ref}}^{19}$ & \nodata & -1 (f) & \nodata & -1 (f) \\
		    & norm$^{20}$ ($\times10^{-3}$) & \nodata & 15$^{+38}_{-4}$ & \nodata & 19.0$\pm0.2$ \\
		    & $\chi^2$/dof & $\frac{216}{224}\approx0.96$ & $\frac{240}{224}\approx1.1$ & $\frac{217}{232}\approx0.94$ & $\frac{257}{230}\approx1.2$ \\ 
			& Flux$_{0.7-50~\rm{keV}}$ (unabsorbed) & $1.3\times10^{-8}$ & $1.5\times10^{-8}$ & $1.3\times10^{-8}$ & $1.5\times10^{-8}$ \\
	\hline 
    \end{tabular}
    \begin{tablenotes}
    \item \textbf{Notes.} $^1$Galactic absorption column density, $^2$ Inner disk temperature, $^4$Photon index, $^5$Electron temperature, $^6$Covering fraction, $^7$Spin of the black hole, $^8$Disk inclination, $^9$Mass of the black hole, $^{10}$Mass accretion rate in the unit of g s$^{-1}$, $^{11}$Distance from the source in the unit of kpc, $^{13}$\texttt{gaussian} line energy, $^{14}$\texttt{gaussian} line width, $^{15}$Emissivity index, $^{16}$Inner disk radius in the unit of $R_{\rm{isco}}$, $^{17}$Log of ionization parameter of the accretion disk, $^{18}$Iron abundance in the unit of Solar abundance, $^{19}$Reflection fraction, $^{3, 12 ,20}$ normalization parameter of the corresponding spectral parameter, $^\ddagger$ values of the errors are trivial and hence are not quoted, f-- indicates the fixed parameters, Flux is in the unit of erg cm$^{-2}$ s$^{-1}$.
    \end{tablenotes}
\end{table*}

In order to calculate the physical parameters such as the mass and spin of the black hole, as well as the mass accretion rate, we replaced the {\tt diskbb} model with the relativistic accretion disk model {\tt kerrbb} \citep{Li et al.2005} in Model-1. Here, we convolved the disk spectrum originating from the {\tt kerrbb} model with {\tt thcomp}, and thus the model setup {\tt constant*tbabs(thcomp*kerrbb+gaussian)}, termed as the Model-2. This model provides a better fit in comparison with the Model-1 with $\chi^2/$dof $= 235.5/237$. In addition, we have followed \citet{Chakraborty et al.2021} and derived Akaike Information Criterion \citep[AIC;][]{Akaike1974} and Bayesian Information Criterion \citep[BIC;][]{Schwarz1978} to show that out of these two models (Model-1 and Model-2) which one represents the data well. The AIC is defined as $\rm{AIC}=-2~ln~\mathcal{L}+P$, where $\mathcal{L}$ is the likelihood of the best-fitting model, P is the number of free parameters. On the other hand, the BIC is defined as $\rm{BIC}=-2~ln~\mathcal{L}+P~ln~(N)$, where N is the total number of data points. Here, the likelihood ($\mathcal{L}$) can be calculated as $-2~\rm{ln}~\mathcal{L}=\chi^2_{\rm{min}}$, where $\chi^2_{\rm{min}}$ is the value of $\chi^2$ for the best-fitting model. The model having the lowest values of AIC and BIC is considered as the most plausible model representing the data. The respective estimated values of AIC and BIC are 356.6 and 405.7 for Model-1, whereas those of Model-2 are found to be 202.6 and 296.1. From these calculated values of AIC and BIC, it is clear that the Model-2 fits the data better than the Model-1. The best-fit spectral parameters obtained using the Model-2 are given in Table 3, and the best-fit model is shown in Figure 9. Here, we fixed the normalization parameter of the {\tt kerrbb} component at unity and the spectral hardening factor at $1.7$ \citep{Shimura and Takahara1995}. We also fixed the distance of the source (D) at 8 kpc, considering the source to be at the Galactic center \citep{Shidatsu et al.2021}. The Galactic absorption column density found from this model is 
$\sim0.3\times10^{22}$ cm$^{-2}$. We found the $90\%$ lower limit on the spin parameter (a) of the black hole to be 0.95. The mass of the black hole is well constrained and found to be $M_{\rm{BH}}=14\pm2$ M$_\odot$. From this model, we found the inclination angle $i=48^{+4}_{-5}$ $^\circ$, and the rate of mass accretion by the compact object from the companion star is $\dot{M}$ $=0.35^{+0.08}_{-0.05}$ $\times 10^{18}$ g s$^{-1}$.
The estimated value of the photon index is $\Gamma$ $\sim2$. Moreover, the temperature of the electron cloud and the covering fraction are found to be $kT_{\rm{e}}$ $\sim17$ keV and $cov_{frac}$ $\sim0.20$, respectively. Here, we also kept the centroid line energy of the iron line fixed at $6.5$ keV and  line width free to vary. The best-fit value of the line width and the estimated EW of the iron line using the Model-2 are found to be $1.4\pm0.2$ keV and $0.19^{+0.04}_{-0.05}$ keV, respectively.

Further, we attempted to model the full reflection spectrum using the relativistic reflection model {\tt relxillCp}. The model {\tt relxillCp} is a part of the blurred reflection model {\tt relxill} \citep{Garcia et al.2014}, which internally includes a physical Comptonization continuum. In this regard, we replaced the {\tt gaussian} component in the Model-2 with {\tt relxillCp}, and thus the full model setup becomes {\tt constant*tbabs*(thcomp*kerrbb+relxillCp)}, which is referred as Model-3. Here, the X-ray continuum is fitted with {\tt kerrbb}, whereas {\tt relxillCp} is considered only as a reflection component by keeping the $R_{Ref}$ parameter fixed at $-1$. The spin and inclination parameters were tied across {\tt relxillCp} and {\tt kerrbb} and kept them free. In addition, the photon index ($\Gamma$) was tied between {\tt relxillCp} and {\tt thcomp} and also kept free. We considered a single emissivity profile ($\epsilon \varpropto r^{-q}$, where $q$ is the emissivity index) throughout the accretion disk and therefore, tied the break radius with the outer disk radius at $400$r$_g$. We tried to vary $q$ freely but found that it was pegging to the lowest defined value of 3 and hence kept it fixed at 3. We also fixed the value of iron abundance ($\rm{A_{Fe}}$) to the solar value. Since the Model-3 appears to be slightly overfitted with the use of $3\%$ systematic error to both the SXT and LAXPC20 spectral data, we considered $2\%$ systematic for each spectrum for this model. This provides a fit with $\chi^2/$dof $= 256.6/236$. However, no significant changes in the parameter values are found over the change of this systematic error. The best-fit spectral parameters obtained using Model-3 are listed in Table 3, and the best-fit model is depicted in Figure 9. The AIC and BIC values obtained for Model-3 are 268 and 323, respectively. We found the best-fit value of the Galactic absorption column density ($N_H$) to be $\sim0.45\times10^{22} \rm{cm}^{-2}$. The photon index ($\Gamma$), estimated from the {\tt thcomp} is found to be $\sim2.13$. However, the temperature of the electron cloud could not be constrained and its $90\%$ lower limit is found to be 95 keV. The spin and inclination parameters, estimated by considering both the reflection and continuum components are found to be $0.94^{+0.03}_{-0.28}$ and $i=57^{+17}_{-6}$, respectively. Using the Model-3, the estimated mass of the black hole is $M_{\rm{BH}}=10.5\pm2.0$ $M_\odot$. The measured inner radius of the accretion disk from the {\tt relxillCp} is $R_{\rm{in}}=3.0^{+2.2}_{-1.5}$ $R_{\rm{isco}}$. Moreover, we found the ionization parameter of the disk to be high with $log\xi \sim4$.

Since time-dependent QPO properties and a slight change in the HR from seg-1 to seg-2 have been observed in this source (see Figure~1), we extracted both SXT and LAXPC20 spectral data in these two time segments and performed broadband spectral analysis using Model-2 and 3. Similar to the above, we have incorporated a $3\%$ systematic for Model-2 and $2\%$ in Model-3. The spectral modeling indicates that the best-fit model parameters obtained from Model-2 and 3 do not show any significant segment wise variations (see Table~4). Moreover, we have not noticed any clear variation in the unabsorbed source flux derived in the $0.7-50$~keV band over the segments.

\section{Discussion and Concluding Remarks} 
\label{sec:floats}

In this work, we have carried out comprehensive spectral and timing analysis of the newly discovered X-ray transient MAXI~J1803--298 using an AstroSat observation (on MJD 59345) during its outburst in May 2021, when the source was in the HIMS.

We have detected LFQPOs along with a sub-harmonic in the $3-15$ keV band. Both the frequency and the fractional rms amplitude of the QPO are found to be increased in the $15-30$ keV band in comparison to those found in the $3-15$ keV band, implying an energy dependent nature of the QPO (see Table~1). The absence of the sub-harmonic in this band may be due to the low S/N of the data. The shape of the PDS, quality factor and fractional rms amplitude of both the QPOs are found to be similar to the characteristics of the type C QPOs \citep{Casella Belloni and Stella2005, Motta et al.2011, Belloni et al.2020}. As we found an increase in the fractional rms amplitude of the QPO with energy in the HIMS, similar increase in fractional rms amplitude of the LFQPO (at $\sim0.16$ Hz) with energy was also reported in the LHS of the source by \citet{Wang et al.2021} using an Insight-HXMT observation on MJD 59337. However, unlike the energy dependent nature of the QPO observed in this present work, they found the QPO to be energy independent. NICER also observed the source in the LHS on MJD 59336 and detected a LFQPO at $\sim0.13$ Hz, which was found to be drifted to a  higher frequency of $\sim0.23$ Hz in the interval of two days on MJD 59338 \citep{Bult et al.2021}. A LFQPO at a comparatively higher frequency of $\sim0.4$ Hz was also detected by a NuSTAR observation on MJD 59340, when the source was in LHS \citep{Xu and Harrison2021}. Following this, we observed the source on MJD 59345 in the HIMS and detected the QPOs at comparatively higher frequencies of $\sim5.4$ Hz and $\sim6.3$ Hz along with a sub-harmonic at $\sim2.8$ Hz. Again the NICER follow-up observations of MAXI~J1803--298 over the period of MJD 59353 to 59359 detected the type-B QPO at $\sim6$ Hz along with a broad sub-harmonic at $\sim3$ Hz, when the source stepped into the intermediate steep power-law state \citep{Ubach et al.2021}. During this period the QPO frequency moved between $\sim6.5$ Hz and $\sim5$ Hz. It is clear that the QPO moved from the low to high frequency, and also changed its nature as the source made a state transition during this particular outburst. As discussed above, the intermittent presence of the QPOs and the change in their frequencies with time and energy indicate the evolution in the source geometry from the rising phase of the outburst to the decaying phase. Along with the energy dependent nature, the QPO also shows a time-dependent variation. The DPS derived over the full exposure time of this particular AstroSat observation (see Figure~3) suggests that the change in the QPO frequency may have a connection with the slight drop in the count rate observed in the seg-2 of the lightcurve (see Figure~1). This can also be confirmed from Table~2, where the QPO frequency goes up from $\sim5.4$ Hz to $\sim6.4$ Hz and finally goes down to $\sim5.5$ Hz.
Although the periodic dips at the interval of $\sim7$ hrs have been reported in a few follow-up observations during the 2021 outburst \citep{Homan et al.2021, Xu and Harrison2021, Jana et al.2021, Jana et al.2022}, it was not noticed in the time span of our observation.

The study of the energy dependent timing properties in the BHTs provides important insights to differentiate the properties of the intermediate states from those of the soft states. The appearance of the type-C QPOs is one of the most salient features seen in the HIMS, and therefore, the study of these QPOs acts as an important tool to further investigate of characteristics of BHTs in the HIMS \citep{Bu et al.2021}. Although the exact physical origin of these QPOs is still not clear, sufficient indications have been found in support of the geometric origin of the type-C QPOs \citep{Gilfanov2010}. The Lense-Thirring precession of a radially extended region of the hot inner flow in the truncated disk models is usually thought to be responsible for the appearance of the LFQPOs in the PDSs of 
the BHTs, suggesting the geometric origin of the type-C QPOs \citep{Stella and Vietri1998, Ingram et al.2009}. Moreover, a number of studies have been carried out in support of the geometric origin of the type-C QPOs by probing the dependence of the timing properties on orbital inclination angle \citep{Schnittman et al.2006, Motta et al.2015}. As in Figure 5, the fractional rms variability amplitudes of both the QPOs initially increase with photon energy up to $\sim12$ keV, and then become either flat or decrease slightly that may be due to the large variability of the power-law component with respect to the disk emission and reflection component. In this regard, it is to note here that a large number of studies are conducted on the energy dependence of the fractional rms variability amplitude of type-C QPOs for several BHXRBs such as GX~339--4 
\citep{Belloni et al.2011}, Cyg~X--1, XTE~1752--223 \citep{MunozDarias et al.2010}, GRS~1915+105 \citep{Rodriguez et al.2004, Yan et al.2012, Yadav et al.2016b, Rawat et al.2019, Zhang et al.2020, Karpouzas et al.2021}, XTE~J1650--500 \citep{Gierlinski and Zdziarski2005}, XTE~J1859+226 \citep{Casella et al.2004}, H~1743--322 \citep{Li et al.2013a}, XTE~J1550--564 \citep{Li et al.2013b} and MAXI~J1631--479 \citep{Bu et al.2021}, where an increase in the amplitude of the fractional variability up to $\sim10$ keV, and then a flattening or decrease has been found. This increase in the fractional rms of the QPOs with photon energy (positive correlation) is usually found in the intermediate states, and the positive correlation implies the origin of the QPOs from the coronal region \citep[see][]{Li et al.2013a, Li et al.2013b}. Apart from this, \citet{You et al.2018} performed the simulation of the fractional rms amplitudes of the type-C QPOs using the Lense-Thirring precession model and suggested that the flattening above 
$\sim10$ keV is caused by the effect of high orbital inclination angle. \citet{Feng et al.2021} also suggested that the source MAXI~J1803--298 has a high disk orbital inclination angle, $i\sim70^\circ$. The value of disk orbital inclination estimated from the Model-3 (see Table~3) hints that this source may be a highly inclined system. Moreover, a flat and decreasing nature in the fractional rms variability with energy (inverted correlation) of the type-C QPOs is observed in the LHS of some of the BHXRBs \citep{Rodriguez et al.2004, Gierlinski and Zdziarski2005, Sobolewska and Zycki2006, Li et al.2013a, Li et al.2013b, Chand et al.2020}. 
 A few more studies on the energy dependent fractional rms variability above 30 keV have been performed for the BHXRBs GRS~1915+105 \citep{Tomsick and Karret2001}, MAXI~J1535--571 \citep{Huang et al.2018} and MAXI~J1820+070 \citep{Ma et al.2020}, which also indicates the geometric origin of the type-C QPOs.
 
With respect to the $3-5$ keV band, the decreasing trend in the time-lags with energy estimated at the QPOs and the sub-harmonic frequencies (see Figure~7) implies the presence of soft or negative lag, where the hard photons arrive before the soft ones. The presence of soft lag is also supported by frequency-lag spectra, where negative lags are observed at both the QPOs and the sub-harmonic frequencies (see Figure~6). The systematic study of the fifteen BHXRBs using RXTE archival data by \citet{van den Eijnden et al.2017} revealed that the sign of the energy-dependent lags strongly depends on the orbital inclination. Besides, the change in the behavior of the time-lags properties with inclination has been studied by \citet{Dutta and Chakrabarti2016} for the BHXRB sources GX~339--4 and XTE~J1550--564, and they have suggested that the soft or negative lag may appear in the high-inclination sources due to reflection and focusing effects.
Moreover, considering the $3-4$ keV band as the reference band, \citet{Yadav et al.2016b} found a decreasing trend in the time-lags up to $\sim20$ keV, and then an upturn in the lag-energy spectrum of GRS~1915+105 using AstroSat observations. Based on this result, they suggested the precession of the inner disk due to the Lense-Thirring effect to be responsible for the origin of the QPO, and the observed time-lags is due to light travel time effects of the irradiation by the inner disk on the non-precessing outer one. In this case, the upturn in the time-lags above $\sim20$ keV is suggested to be due to the time-delay of the reflected photons with respect to continuum at $\sim20$ keV.

We have modeled the time averaged X-ray energy spectra from both SXT and LAXPC20 jointly using three model combinations from Model-1 to 3 (see Table~3). Based on the statistics, we have found that Model-2 describes the spectra in a better way than the other two models. The photon index ($\Gamma$) obtained from the {\tt thcomp} component from the best-fit model Model-2 indicates that the source was in the HIMS during this particular observation. Moreover, its values obtained from the other two models are found to be nearly similar. It is worth mentioning here that \citet{Jana et al.2022} also observed the same source immediately after our observation and similarly found the source to be in the HIMS.
The Galactic absorption column density ($N_H$) derived from Model-1 to Model-3 
is found to be in the range of $\sim(0.22-0.45)\times10^{22}$ cm$^{-2}$. These values are almost consistent with the values reported by \citet{Bult et al.2021}, \citet{Feng et al.2021} and \citet{Jana et al.2022}. However, the larger value of the absorption column density obtained from Model-3 may be due to the soft excess introduced by {\tt relxillCp} model at low energies. The inner disk temperature ($kT_{in}$) obtained from the {\tt diskbb} model suggests the accretion disk to be hot. The line width of the iron line estimated from Model-1 and 2 is similar within errors (See Table~3) and the presence of such broad iron line suggests its origin in the innermost regions of the accretion disk, where the relativistic effects play the role to distort the shape of the line \citep{Fabian et al.2000}. In addition, we find that the X-ray spectrum is dominated by the power-law component with a power-law fraction of $\sim60\%$, whereas the accretion disk contributes only $\sim30\%$. The slight discrepancy in the values of the same parameters derived from the chosen models may arise from the different assumptions made in the model components. In this regard, it is worth mentioning here that the \texttt{kerrbb} represents a relativistic accretion disk, which considers the disk to be extended close to the ISCO. Moreover, \texttt{kerrbb} also accounts for the effect of self irradiation and either zero or non-zero torque at the inner boundary of the disk \citep{Li et al.2005}. On the other hand, {\tt diskbb} neglects the relativistic effects and represents a thin Newtonian disk \citep{Mitsuda et al.1984, Makishima et al.1986}. Apart from this, the {\tt diskbb} does not consider the effect of irradiation and zero-torque at the inner boundary of the disk \citep{Stiele and Kong2017}. We have found a black hole mass of 14$\pm2$ $M_\odot$, spin ($a$) $>0.95$ ($90\%$ confidence level) and the inclination angle $(i)$ of 48$^{+4}_{-5}$ based on the {\tt kerrbb} model, which only uses the disk continuum. 
The similar parameters have also been estimated from Model-3 by considering both the disk continuum ({\tt kerrbb}) and the reflection spectrum ({\tt relxillCp}). We find that the parameters such as mass and spin of the black hole and the disk inclination angle are consistent within errors with those values estimated from Model-2. From the Model-2 and 3, the estimated black hole mass ranges from $\sim 8.5-16$ $M_\odot$, and the spin parameter is $\gtrsim0.7$.
In addition, the disk inclination angle derived from the above mentioned two models ranges from $\sim43-74$ $^\circ$. These values are consistent with the recent works by \citet{Feng et al.2021} and \citet{Jana et al.2022}, where the authors also detected the presence of significant periodic dips in the lightcurves. We note that the estimated mass and spin of the black hole are completely based on the X-ray spectroscopy and depend on the distance of the source, which is assumed to be 8 kpc. 
The measured radius of the inner accretion disk (see Table~3) by modeling the reflection features suggests that the accretion disk is extended close to the ISCO. Small inner disk radius in the hard-intermediate state for MAXI~J1535--571 and GX~339--4 is also been reported by \citep{Sridhar et al.2019, Sridhar et al.2020} using RXTE and AstroSat observations. The disk is found to be highly ionized. 
To understand the dependency of the mass and spin parameter of the black hole on the distance to the source, we have varied the distance in the range of $6-10$ kpc in both Model-2 and 3. However, we find that the best-fit values of the black hole mass and spin obtained from the Model-2 and 3 are consistent within errors with those derived by assuming the distance to be 8 kpc for the respective models. Similarly, no significant variation is also observed in the mass accretion rate.
 Moreover, unlike the time dependent behavior of the QPO properties, we do not find any significant changes in the spectral parameters over the time (see Table~4). All the above results obtained through the temporal and spectral analysis indicate that the newly discovered X-ray transient source MAXI~J1803--298 is possibly a black hole X-ray binary hosting a stellar mass Kerr black hole.


\acknowledgments

We thank the anonymous referee for the useful comments and suggestions that have improved the quality of the paper.
The authors acknowledge the financial support of ISRO under AstroSat archival Data utilization program (No: DS-2B-13013(2)/8/2019-Sec.2). This publication uses data from the AstroSat mission of the Indian Space Research Organisation (ISRO), archived at the Indian Space Science Data Centre (ISSDC). This work has used the data from SXT and LAXPC instruments onboard AstroSat. LAXPC data were processed by the Payload Operation Center (POC) at TIFR, Mumbai. This work has been performed utilizing the calibration data-bases and auxiliary analysis tools developed, maintained and distributed by the AstroSat-SXT team with members from various institutions in India and abroad, and the SXT POC at the TIFR, Mumbai \url{(https://www.tifr.res.in/~astrosat_sxt/index.html)}. SXT data were processed and verified by the SXT POC. S.C expresses his sincere thanks to David P. Huenemoerder for his help.
VKA thanks GH, SAG; DD, PDMSA and Director, URSC for encouragement and continuous support to carry out this research. P.T. expresses his sincere thanks to the Inter-University Centre for Astronomy and Astrophysics (IUCAA), Pune, India, for granting supports through the IUCAA associateship program. S.C. is also very much grateful to IUCAA, Pune, India, for providing support and local hospitality during his frequent visits.

%

\vspace{5mm}
\facilities{AstroSat(SXT and LAXPC)}


\software{LaxpcSoft, SXT Software, HEASoft (V. 6.26), ISIS, Julia \citep{Bezanson et al.2017}, XSPEC \citep[12.10.1;][]{Arnaud1996}, GNUPLOT}

\end{document}